\documentclass[10pt, twocolumn]{IEEEtran}

\usepackage{bm}
\usepackage{url}
\usepackage{bbm}
\usepackage{cite}
\usepackage{array}
\usepackage{ifthen}
\usepackage{xspace}
\usepackage{dsfont}
\usepackage{siunitx}
\usepackage{balance}
\usepackage{amsmath}
\usepackage{amssymb}
\usepackage{multicol}
\usepackage{amsfonts}
\usepackage{mathrsfs}
\usepackage{booktabs}
\usepackage{graphicx}
\usepackage{caption}
\usepackage{setspace}
\usepackage{hyperref}
\usepackage{makecell}
\usepackage{footnote}
\usepackage{verbatim}
\usepackage{algorithm}
\usepackage{subcaption}
\usepackage{glossaries}
\usepackage{soul,xcolor}
\usepackage[T1]{fontenc}
\usepackage{algpseudocode}
\usepackage{algcompatible}
\usepackage[normalem]{ulem}
\usepackage{multirow,enumitem}

\title{Propagation Measurements and Analyses at \SI{28}{\giga\hertz} via an Autonomous Beam-Steering Platform}
\author{Bharath Keshavamurthy\IEEEauthorrefmark{1}, Yaguang Zhang\IEEEauthorrefmark{2}, Christopher R. Anderson\IEEEauthorrefmark{3},\\Nicol\`{o} Michelusi\IEEEauthorrefmark{1}, James V. Krogmeier\IEEEauthorrefmark{2}, and David J. Love\IEEEauthorrefmark{2}
\thanks{\IEEEauthorrefmark{1}Electrical, Computer and Energy Engineering, Arizona State University.}
\thanks{\IEEEauthorrefmark{2}Electrical and Computer Engineering, Purdue University.}
\thanks{\IEEEauthorrefmark{3}Electrical Engineering, United States Naval Academy.}
\thanks{A summary of this research was presented at URSI NRSM 2022~\cite{NRSM}.}
\thanks{Source code available on \href{https://github.com/bharathkeshavamurthy/SPAVE-28G.git}{GitHub}\cite{SPAVE-28G-Software}. Dataset available on \href{https://doi.org/10.5281/zenodo.7178597}{Zenodo}\cite{SPAVE-28G-Dataset}.}
\thanks{Research funded by NSF under grants CNS-1642982 and CNS-2129615.}
\vspace{-12mm}
}

\begin{document}
\bstctlcite{IEEEexample:BSTcontrol}

\maketitle
\thispagestyle{empty}
\pagestyle{empty}
\setulcolor{red}
\setul{red}{2pt}
\setstcolor{red}
\vspace{-12mm}

% Abstract
\begin{abstract}
This paper details the design of an autonomous alignment and tracking platform to mechanically steer directional horn antennas in a sliding correlator channel sounder setup for \SI{28}{\giga\hertz} V$\bm{2}$X propagation modeling. A pan-and-tilt subsystem facilitates uninhibited rotational mobility along the yaw and pitch axes, driven by open-loop servo units and orchestrated via inertial motion controllers. A geo-positioning subsystem augmented in accuracy by real-time kinematics enables navigation events to be shared between a transmitter and receiver over an Apache Kafka messaging middleware framework with fault tolerance. Herein, our system demonstrates a $\bm{3}$D geo-positioning accuracy of \SI{17}{\centi\meter}, an average principal axes positioning accuracy of \SI{1.1}{\degree}, and an average tracking response time of \SI{27.8}{\milli\second}. Crucially, fully autonomous antenna alignment and tracking facilitates continuous series of measurements, a unique yet critical necessity for millimeter wave channel modeling in vehicular networks. The power-delay profiles, collected along routes spanning urban and suburban neighborhoods on the NSF POWDER testbed, are used in pathloss evaluations involving the $\bm{3}$GPP TR$\bm{38.901}$ and ITU-R M$\bm{.2135}$ standards. Empirically, we demonstrate that these models fail to accurately capture the \SI{28}{\giga\hertz} pathloss behavior in urban foliage and suburban radio environments. In addition to RMS direction-spread analyses for angles-of-arrival via the SAGE algorithm, we perform signal decoherence studies wherein we derive exponential models for the spatial/angular autocorrelation coefficient under distance and alignment effects.
\end{abstract}
\vspace{-10mm}

% Introduction, Literature survey, and Novelties
\section{Introduction}\label{S1}
The widespread adoption and recent acceleration in the deployment of $5$G networks have resulted in an increased strain on the mid-band spectrum (\SIrange{4}{8}{\giga\hertz}), thereby causing providers around the world to shift their focus onto the millimeter wave bands (mmWave: \SIrange{30}{300}{\giga\hertz})~\cite{Commercial}. With mmWave deployments, both consumers and businesses can exploit the relatively higher bandwidth over extant mid-band networks: consumers can experience better quality-of-service vis-\`{a}-vis data rates at their user terminals, while businesses can leverage reliable service guarantees to optimize operations~\cite{Rappaport}. 

While mmWave networks promise quality-of-service gains, the signals suffer from poor propagation characteristics, i.e., increased atmospheric attenuation and absorption~\cite{Rappaport}. Ergo, there have been concerted efforts from both academia and industry to develop well-rounded mmWave channel models for indoor and outdoor radio environments~\cite{Foliage, Indoor60G, NISTModeling}. Current efforts comprise a wide array of measurement campaigns and subsequent analyses~\cite{Agile-Link, Harvard, Purdue, Foliage, MolischSpatialOutdoor, Outdoor28G, Commercial, MacCartneySpatialStatistics}. These works suffer from poor system design approaches that introduce drawbacks vis-\`{a}-vis cost (expensive phased-arrays~\cite{Agile-Link}), computational complexity (exhaustive signal sampling~\cite{Agile-Link}), and ease of operations (inflexible alignment~\cite{Foliage}); fail to address diversity in transmitter (Tx) and receiver (Rx) deployment~\cite{Purdue, MolischSpatialOutdoor}; fail to empirically validate standardized pathloss models in diverse propagation conditions~\cite{Outdoor28G, Commercial}; and fail to analyze signal spatial consistency behavior under continuously varying Tx-Rx distance and alignment accuracy effects~\cite{Indoor60G, MolischSpatialOutdoor, MacCartneySpatialStatistics}. To address these limitations, this paper describes the design of a sliding correlator channel sounder~\cite{Purdue} in conjunction with a fully autonomous robotic beam-steering platform, employed in a V$2$X propagation modeling campaign on the NSF POWDER testbed~\cite{POWDER} along urban and suburban routes at \SI{28}{\giga\hertz}.

The novel design of our measurement platform enables several key features essential to mmWave V2X propagation modeling. First, our system facilitates greater deployment diversity along routes in both urban and suburban settings (location diversity); allows operations with manual, full-, and semi-autonomous alignment (beam-steering diversity); and enables the use of different mobile mounts (vehicle/cart, velocity diversity). In contrast, the \SI{28}{\giga\hertz} measurement campaign in~\cite{Purdue} centers around a manual antenna alignment platform in suburban neighborhoods under semi-stationary settings only; the sampling activities in~\cite{Harvard} are constrained to indoor environments and the underlying beam-steering platform also requires manual operation; and~\cite{MolischSpatialOutdoor, Outdoor28G} outline semi-stationary approaches restricted to urban deployments. Second, while electronic beam-alignment systems~\cite{Agile-Link} involving phased-array antenna modules offer increased flexibility in side-lobe \& beam-width control and demonstrate faster average tracking response times relative to mechanical fixed-beam steering, they constitute exhaustive beam-search strategies and their design necessitates expensive hardware to host inherently resource-heavy algorithms. Additionally, our work evaluates the signal pathloss behavior, computed from the collected measurements, against the $3$GPP TR$38.901$ and ITU-R M$.2135$ large-scale macro-cellular pathloss models~\cite{MacCartneyModelsOverview}. These comparisons enable the user to gain novel insights on adapting them to rapidly evolving mmWave V$2$V and V$2$I scenarios. Lastly, while~\cite{MolischSpatialOutdoor, MacCartneySpatialStatistics} conduct spatial consistency analyses of mmWave signals vis-\`{a}-vis Tx-Rx distance, they fail to investigate signal decorrelation behavior under variations in Tx-Rx alignment accuracy. With a fully autonomous alignment and tracking platform that aids in the continual series of measurements along unplanned vehicular routes onsite, our design facilitates signal decoherence studies under both distance and alignment effects. The insights gathered from these studies on mmWave signal propagation in vehicular settings can be leveraged in the efficient deployment of $5$G$+$ networks along interstates and in furthering spatial consistency research on V$2$V and V$2$I communications. A glossary of the notations, standards, and protocols referenced in this paper is provided in Table~\ref{T1}.

\indent{The} rest of the paper is structured as: Sec.~\ref{S2} elucidates the end-to-end design of our autonomous robotic beam-steering platform; Sec.~\ref{S3} details our measurement and post-processing activities; Sec.~\ref{S4} describes our numerical evaluations and the insights gained from pathloss and spatial consistency studies; and finally, Sec.~\ref{S5} outlines our concluding remarks.
\vspace{-3mm}

% System design description
\section{Measurement System: Design Description}\label{S2}
\begin{figure*} [t]
    \centering
    \includegraphics[width=1.02\textwidth]{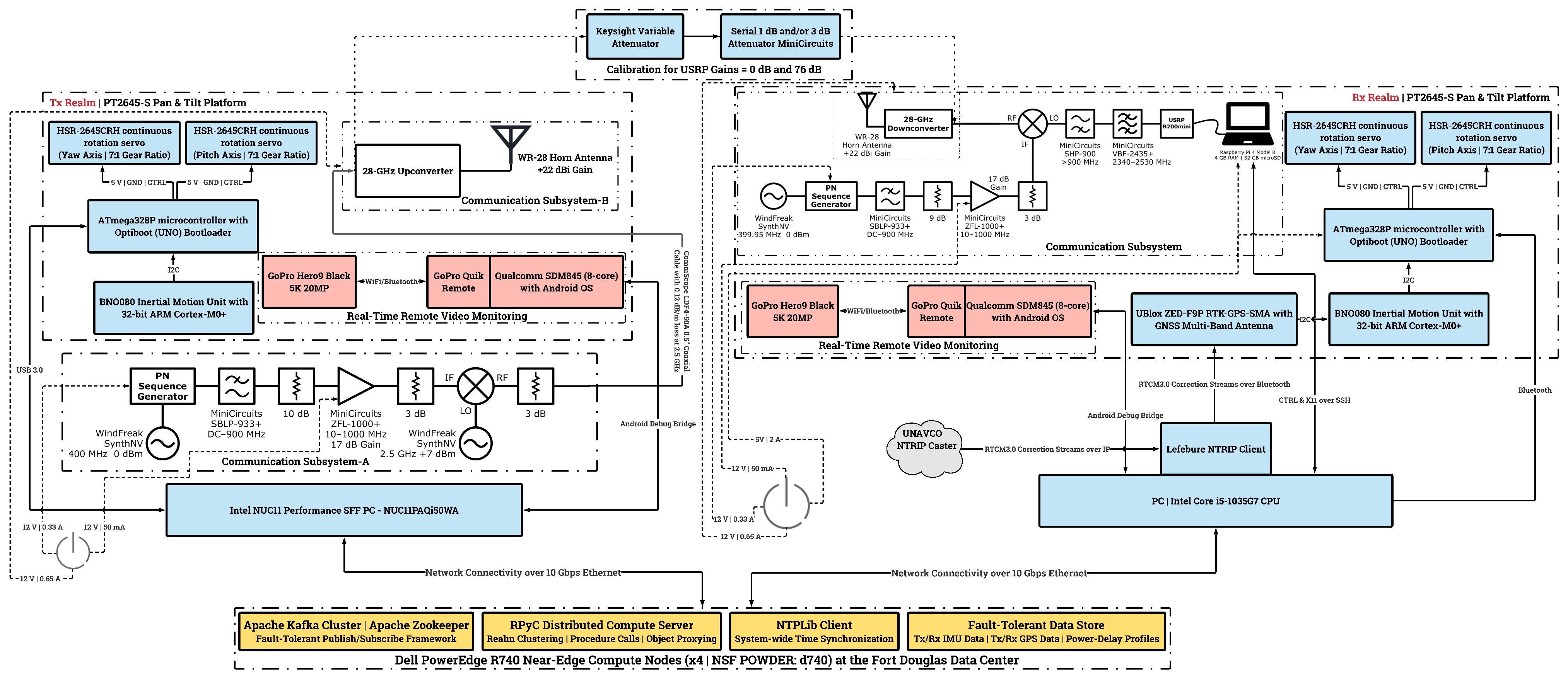}
    \vspace{-6mm}
    \caption{The architecture of our autonomous antenna alignment \& tracking platform with a sliding correlator channel sounder.}
    \label{F1}
    \vspace{-7mm}
\end{figure*}

With the objective of facilitating uninterrupted measurement operations for \SI{28}{\giga\hertz} propagation modeling along unplanned routes in V$2$X settings, our measurement campaign on the NSF POWDER testbed~\cite{POWDER} involved a sliding correlator channel sounder~\cite{Purdue} with directional horn antennas in conjunction with a fully autonomous mechanical beam-steering platform for continual antenna alignment and tracking~\cite{NRSM}. Specifically, under V$2$I evaluations, with a rooftop mounted Tx and a mobile Rx traversing unplanned vehicular routes onsite, this design enables the logging of geo-positioning data (i.e., GPS coordinates, speed, acceleration, and heading), alignment angles, and power-delay profile samples. With the system architecture shown in Fig.~\ref{F1}, in this section, we first discuss our channel sounder design and then describe the development of our autonomous alignment and tracking platform.

\begin{table} [tb]
	\centering
	\scriptsize
	\begin{tabular}{|l||l|}
		\hline
		V$2$X & Vehicle-to-Vehicle (V$2$V) or Vehicle-to-Infrastructure (V$2$I)\\
		\hline
		$3$GPP TR38.901 UMa & $3$rd Generation Partnership Project (Urban Macrocells)\\
		\hline
		ITU-R M$.2135$ UMa & International Telecommunication Union (Urban Macrocells)\\
		\hline
		SAGE, AoA & Space Alternating Expectation Maximization, Angle of Arrival\\
		\hline
		HPBW, SDR & Half-Power Beam-Width, Software Defined Radio\\
		\hline
		SSD, SBC & Solid State Drive, Single Board Computer\\
		\hline
		PWM & Pulse Width Modulation (digital control of servos)\\
		\hline
		GNSS, GPS & Global Navigation Satellite System, Global Positioning System\\
		\hline
		RTCM & Radio Technical Commission for Maritime Services\\
		\hline
		RTK & Real-Time Kinematics (GPS corrections)\\
		\hline
		NMEA-0183 & National Marine Electronics Association (data specification)\\
		\hline
		NTRIP & Networked Transport of RTCM over Internet Protocol\\
		\hline
		UNAVCO & University NAVstar COnsortium (GNSS data provisioning)\\
		\hline
		I$2$C & Inter Integrated Circuit (serial communication bus)\\
		\hline
		NTP & Network Time Protocol (timing synchronization)\\
		\hline
	\end{tabular}
	\vspace{-1mm}
	\caption{Glossary of Notations, Standards, and Protocols}
	\label{T1}
\end{table}

\noindent{\textbf{Channel Sounder}}: The measurement system employed a custom broadband sliding correlator channel sounder at both the Tx and the Rx~\cite{Purdue}, each equipped with a Pseudorandom Noise (PN) sequence generator module producing the required known apriori signal for time-dilated cross-correlation studies, with the Rx module clocked at a slightly lower rate than the Tx; up-/down-converter to transition between the \SI{2.5}{\giga\hertz} and \SI{28}{\giga\hertz} regimes; a vertically polarized WR-28 directional horn antenna; and other commercially available components. This setup is depicted by the schematics in Fig.~\ref{F1} and is implemented by the circuits shown in Fig.~\ref{F2}. The operational specifications of the sounder are listed in Table~\ref{T2}, as detailed also in~\cite{Purdue}. At the Rx, complex-\SI{64}{} I/Q power-delay profiles are recorded onboard an SSD storage drive by a GNURadio sink on a Raspberry Pi SBC via a USRP B$200$mini SDR.

\noindent{\textbf{Alignment \& Tracking}}: First, to enable uninhibited rotational mobility for alignment and tracking in the horizontal and the vertical planes, as shown in Fig.~\ref{F1}, at both the Tx and the Rx, the WR-$28$ antenna is mounted on a PT-$2645$S open-loop pan-and-tilt unit, driven by $2{\times}$ HSR-$2645$CRH continuous rotation servos, each actuating either yaw (horizontal) or pitch (vertical) alignment. These servos are controlled via PWM signals from an ATMega$328$P microcontroller with the angular position (absolute and relative) feedback provided by a BNO$080$ inertial motion unit. This principal axes positioning subsystem demonstrates an average accuracy of \SI{1.1}{\degree} across all fine- \& coarse-grained yaw and pitch movements. Next, for seamless operations in V$2$X scenarios, the alignment platform is augmented with a geo-positioning subsystem consisting of a UBlox ZED-F$9$P GPS unit (with a GNSS multi-band antenna) wherein the positioning accuracy is enhanced by RTCM v$3.0$ RTK correction streams over NTRIP from a UNAVCO caster via a Lefebure client. Demonstrating an average $3$D accuracy of \SI{17}{\centi\meter}, the relevant data members captured by this geo-positioning unit---namely, the coordinate (latitude, longitude, and ellipsoidal altitude), the horizontal speed \& acceleration, and the heading, are communicated to the microcontroller as NMEA-0183 messages over an I2C serial peripheral bus.

\begin{table} [tb]
	\centering
	\scriptsize
	\begin{tabular}{|l||l|}
		\hline
		Carrier Frequency & \SI{28}{\giga\hertz}\\
		\hline
		PN Chip Sequence Length & \SI{2047}{}\\
		\hline
		RF Bandwidth & \SI{800}{\mega\hertz}\\
		\hline
		Tx Chip Rate & \SI{400}{\mega{cps}}\\
		\hline
		Temporal Resolution & \SI{2.5}{\nano\second}\\
		\hline
		Rx Chip Rate & \SI{399.95}{\mega{cps}}\\
		\hline
		Tx Power & \SI{23}{\deci\bel{m}}\\
		\hline
		Tx/Rx Antenna Gain & \SI{22}{\deci\bel{i}}\\
		\hline
		Nominal Tx/Rx Antenna HPBW & \SI{15}{\degree}\\
		\hline
		Measured Tx/Rx Azimuth HPBW & \SI{10.1}{\degree}\\
		\hline
		Measured Tx/Rx Elevation HPBW & \SI{11.5}{\degree}\\
		\hline
		Maximum Measurable Pathloss & \SI{182}{\decibel}\\
		\hline
		GNURadio Sink Center Frequency & \SI{2.5}{\giga\hertz}\\
		\hline
		USRP Gain & \SI{76}{\decibel}\\
		\hline
		USRP Sampling Rate & \SI{2}{\mega{sps}}\\
		\hline
	\end{tabular}
	\vspace{-0.7mm}
	\caption{Sliding Correlator Channel Sounder Specifications}
	\label{T2}
\end{table}

Since our measurement system revolves around a decoupled design with the alignment and tracking platform replicated at both the Tx and the Rx, a centralized nerve-center handles asynchronous module registration \& de-registration via RPyC object proxying, global timing synchronization via NTP, and coordination between the Tx and Rx over fault-tolerant Apache Kafka messaging middleware. With Apache Zookeeper broker management, the samples generated by the principal axes positioning and geo-positioning subsystems are shared over Kafka message queues: the Tx subscribes to the alignment and geo-location messages published by the Rx, and vice-versa, resulting in a scalable event-driven modular architecture. Corroborated both onsite and in the laboratory, this publish-subscribe framework facilitates an average beam-steering response time of \SI{27.8}{\milli\second}, evaluated over \SI{12870}{} interactions. With system monitoring provided over an Android debug bridge and system troubleshooting enabled via serial communication interfaces, our platform demonstrates remote orchestration capabilities: a critical necessity for V$2$X propagation modeling. To augment these remote monitoring and troubleshooting features further, a GNURadio Qt GUI time-sink (with dynamic trigger levels) allows for real-time visualization of the recorded power-delay profiles over an ad-hoc WLAN with the Raspberry Pi SBC; additionally, via the Plotly Dash and MapBox APIs, the Tx and Rx geo-locations are annotated with their relative alignment accuracies and visualized in real-time at a control terminal for onsite validation of the routes traversed on NSF POWDER.
\vspace{-3.3mm}

% Measurement campaign description and Data post-processing procedural explanation
\section{Measurements \& Post-Processing}\label{S3}
In this section, we discuss the operations involved in our \SI{28}{\giga\hertz} V$2$X measurement campaign on the NSF POWDER testbed. First, we describe the system calibration process; next, we outline the onsite system deployment procedure; and finally, we detail the post-processing steps involved in setting up the power-delay profiles recorded at the Rx for pathloss evaluations, multipath component extraction and parameter estimation via SAGE, and spatial consistency analyses.

\begin{figure*}[t]
    \centering
    \begin{tabular}{cc}
        \begin{minipage}{0.522\linewidth} 
        	\centering
            \includegraphics[width=0.8\linewidth]{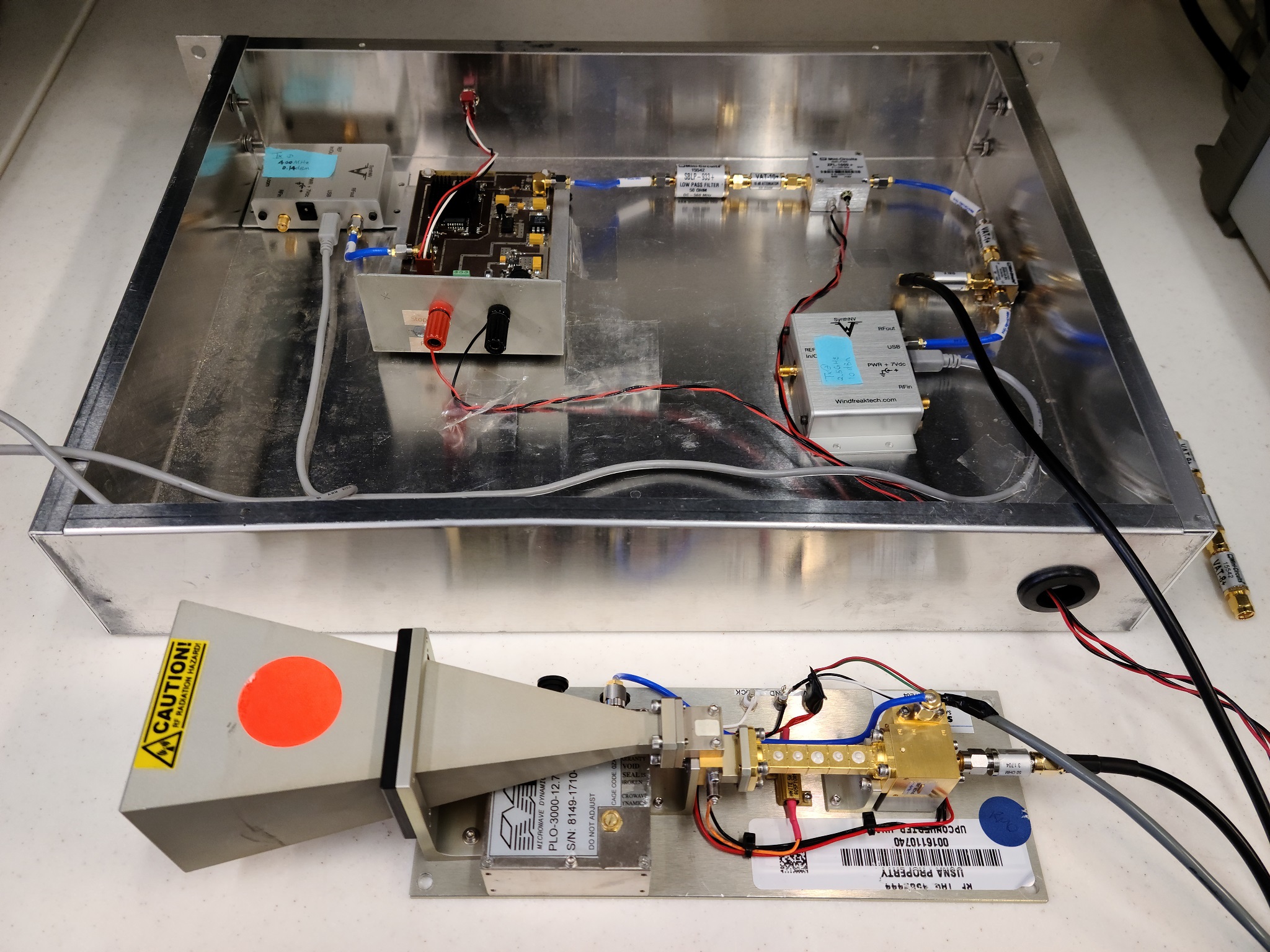}
            \\ [0.55ex]
            \centering
            \includegraphics[width=0.8\linewidth]{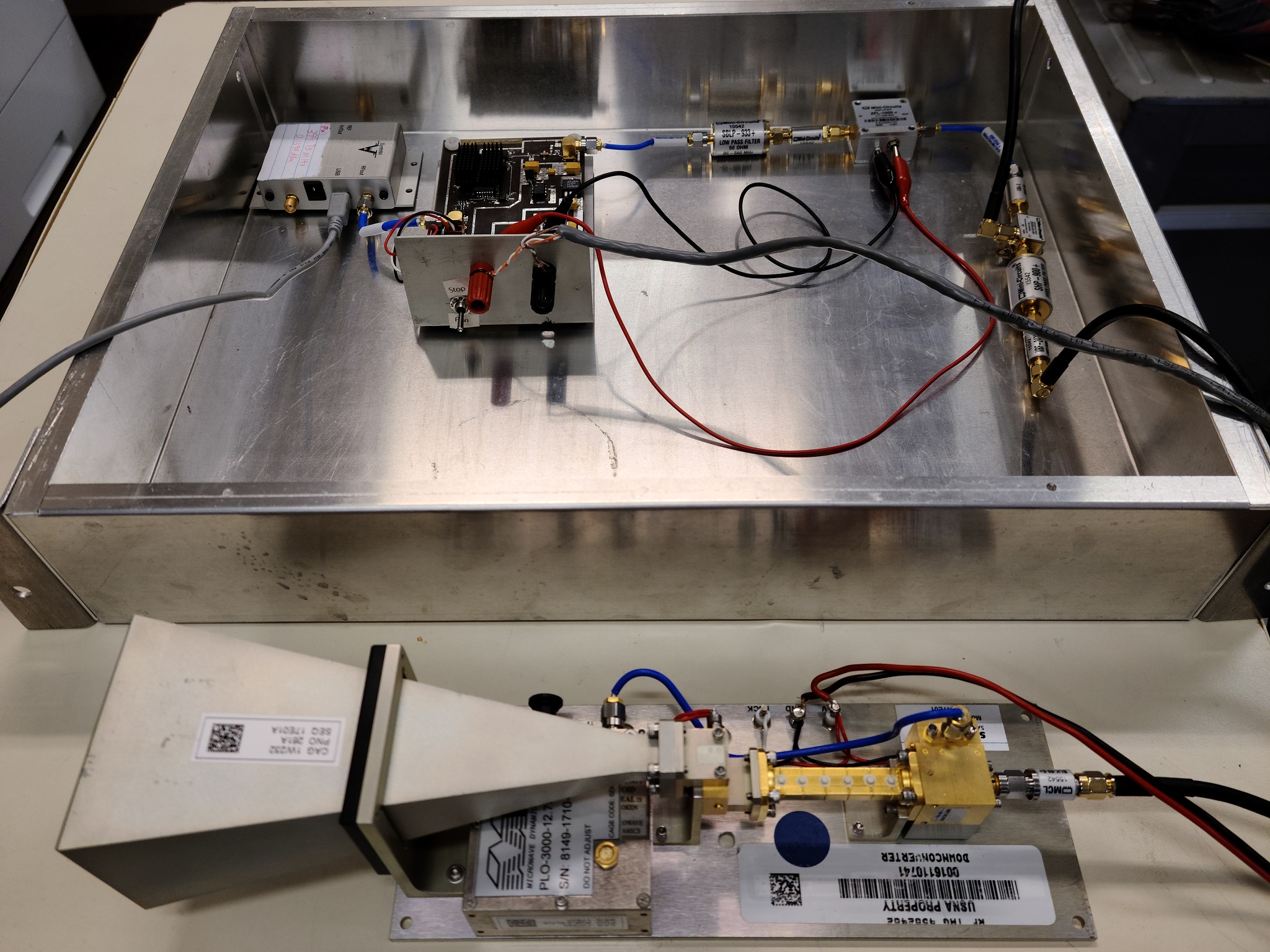}
        \end{minipage}&
        \hspace{-20mm}
        \begin{minipage}{0.478\linewidth}
        	\centering
            \includegraphics[width=0.8\linewidth]{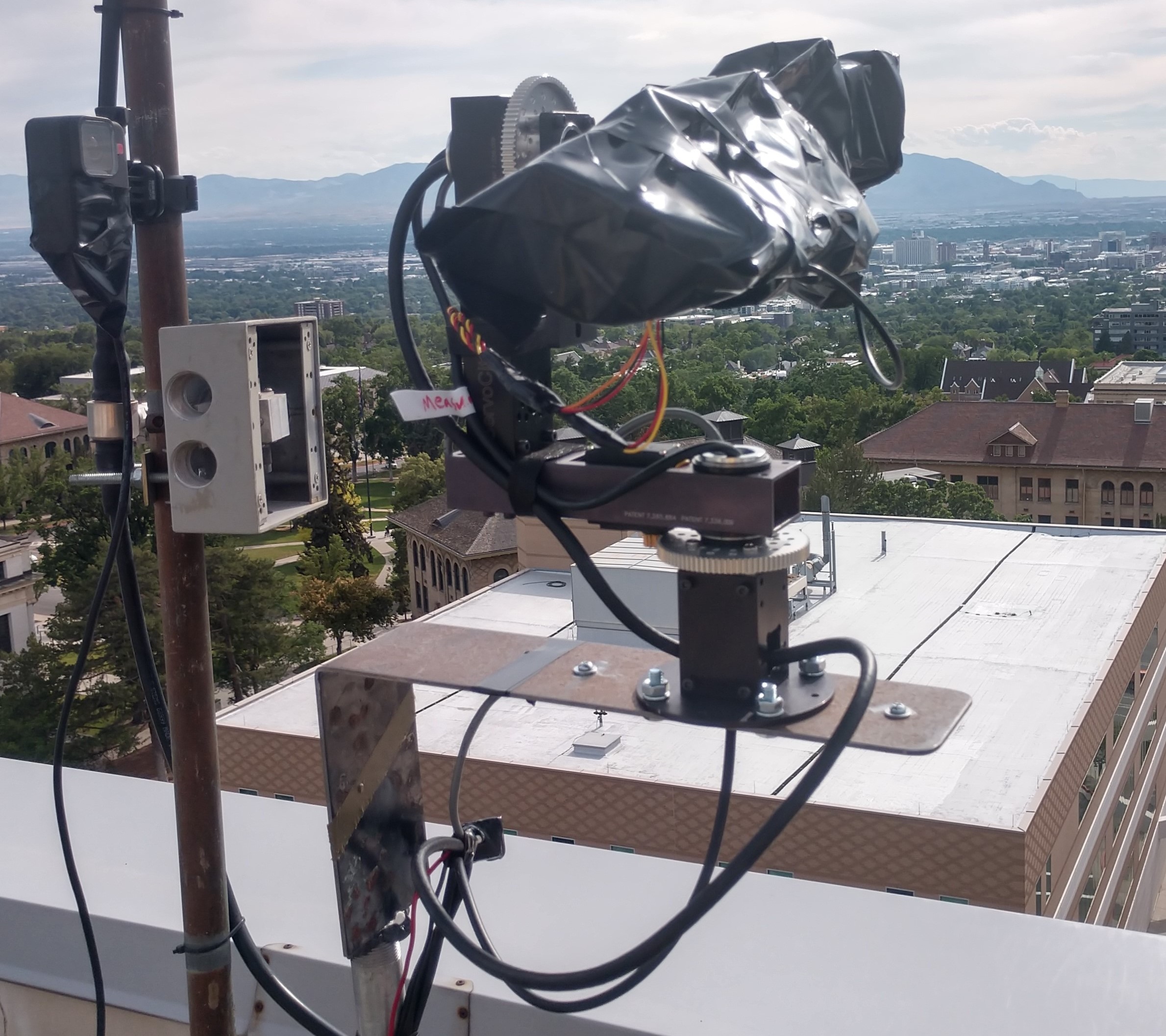}
            \\ [0.55ex]
            \centering
            \includegraphics[width=0.8\linewidth]{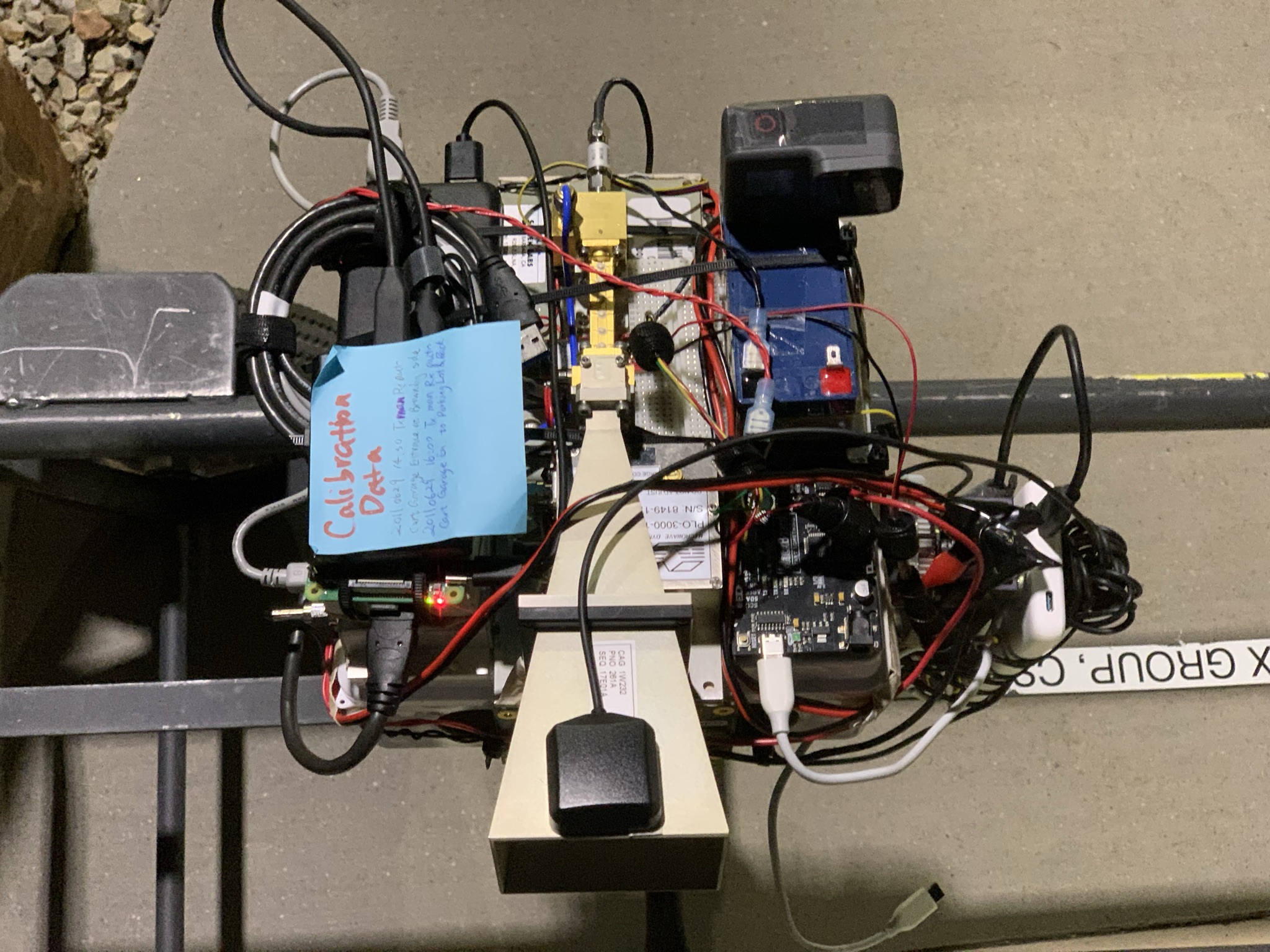}
        \end{minipage}
    \end{tabular}
    \caption{Clockwise from top-left: The Tx circuit with a \SI{28}{\giga\hertz} up-converter, a WR-$28$ directional horn antenna, and other commercially available parts (a); the deployment of the Tx atop the William Browning building: the sounder circuits are housed in a climate-controlled enclosure with the antenna mounted on the pan-and-tilt platform (b); the deployment of the Rx mounted on a push-cart (or a minivan) and pushed (or driven) around onsite (c); the Rx circuit with a \SI{28}{\giga\hertz} down-converter, a WR-$28$ directional horn antenna, USRP B$200$mini SDR, Raspberry Pi SBC, and other commercially available parts (d).}
    \label{F2}
    \vspace{-7mm}
\end{figure*}

\noindent{\textbf{Pre-deployment Calibration}}: After the Tx and Rx circuits for the sliding correlator channel sounder have been implemented as illustrated in Figs.~\ref{F2}(a) and~\ref{F2}(d), a calibration procedure is carried out onsite to map the power calculated from the power-delay profiles recorded by the USRP to reference measured power levels. Calibrating the measurement system before deployment ensures accurate Rx power calculations in the presence of imperfect circuit components, e.g., the Commscope LDF$4$-$50$A \SI{0.5}{{"}} coaxial cables employed at the Tx exhibit losses of up to \SI{0.12}{\deci\bel\per\meter} at \SI{2.5}{\giga\hertz}. Under \SI{0}{\deci\bel} and \SI{76}{\deci\bel} USRP gains, using a Keysight variable attenuator, the recorded power-delay profiles are processed to determine the calculated power values mapped to the reference power levels: the results of this procedure are studied in Sec.~\ref{S4}.

\noindent{\textbf{NSF POWDER Deployment}}: As described in Sec.~\ref{S2}, our measurement system assembly constitutes an autonomous beam-steering controller replicated at both the Tx and the Rx, their respective sounder circuits, and a centralized nerve center for aggregation (RPyC), synchronization (NTP), and coordination (Kafka-Zookeeper). On the NSF POWDER testbed, the nerve center is deployed on a high-availability cluster of four Dell R$740$ compute nodes at the Fort Douglas datacenter, with fault tolerance being a key feature to ensure storage redundancy for the recorded data. As depicted in Fig.~\ref{F2}(b), the Tx is mounted on a building rooftop; while, as shown in Fig.~\ref{F2}(c), the Rx is mounted on a van (or a push-cart) that is driven (or pushed) along unplanned routes onsite. Remote monitoring and troubleshooting is provided for validation of geo-positioning, alignment, and power-delay profile samples. The goal of this measurement campaign was to obtain a reasonably large dataset of site-specific measurements for evaluating the propagation characteristics of \SI{28}{\giga\hertz} signals in vehicular communication settings. Thus, our propagation modeling activities included V$2$I measurements under manual, semi-, and fully-autonomous alignment operations traversing nine routes spanning urban and suburban neighborhoods\footnote{Although our platform is capable of double-directional measurements and facilitates easy scalability to MIMO settings, in this campaign, we focus only on \emph{beam-steered measurements} for spatial consistency evaluations.}.

\noindent{\textbf{Post-Processing}}: Using GNURadio utilities, the metadata file corresponding to the route-specific power-delay profile records at the Rx is parsed to extract timestamp information, which is then associated with the geo-positioning and alignment logs at both the Tx and the Rx. The samples in each synchronized power-delay profile segment undergo pre-filtering via a low-pass filter (SciPy FIR implementation), time-windowing, and noise-elimination (via custom peak-search and thresholding heuristics). Coupled with transmission power and antenna gain values, the received power levels obtained from these processed samples allow the visualization of pathloss maps on the Google Maps API (rendered via the Bokeh toolbox), and the evaluation of pathloss behavior as a function of Tx-Rx distance, with validations against the $3$GPP TR$38.901$ and ITU-R M$.2135$ standards~\cite{MacCartneyModelsOverview}. The SAGE algorithm~\cite{SAGE} is used to extract multipath parameters, which facilitates RMS AoA direction spread studies~\cite{Indoor60G}. Under Tx-Rx distance and Tx-Rx alignment variations, we probe signal decoherence patterns via the spatial/angular autocorrelation coefficient~\cite{MacCartneySpatialStatistics}.
\vspace{-8mm}

% Simulations
\section{Numerical Evaluations}\label{S4}
\begin{figure*} [t]
     \centering
     \begin{subfigure}{0.368\linewidth}
         \centering
         \includegraphics[width=1.0\linewidth]{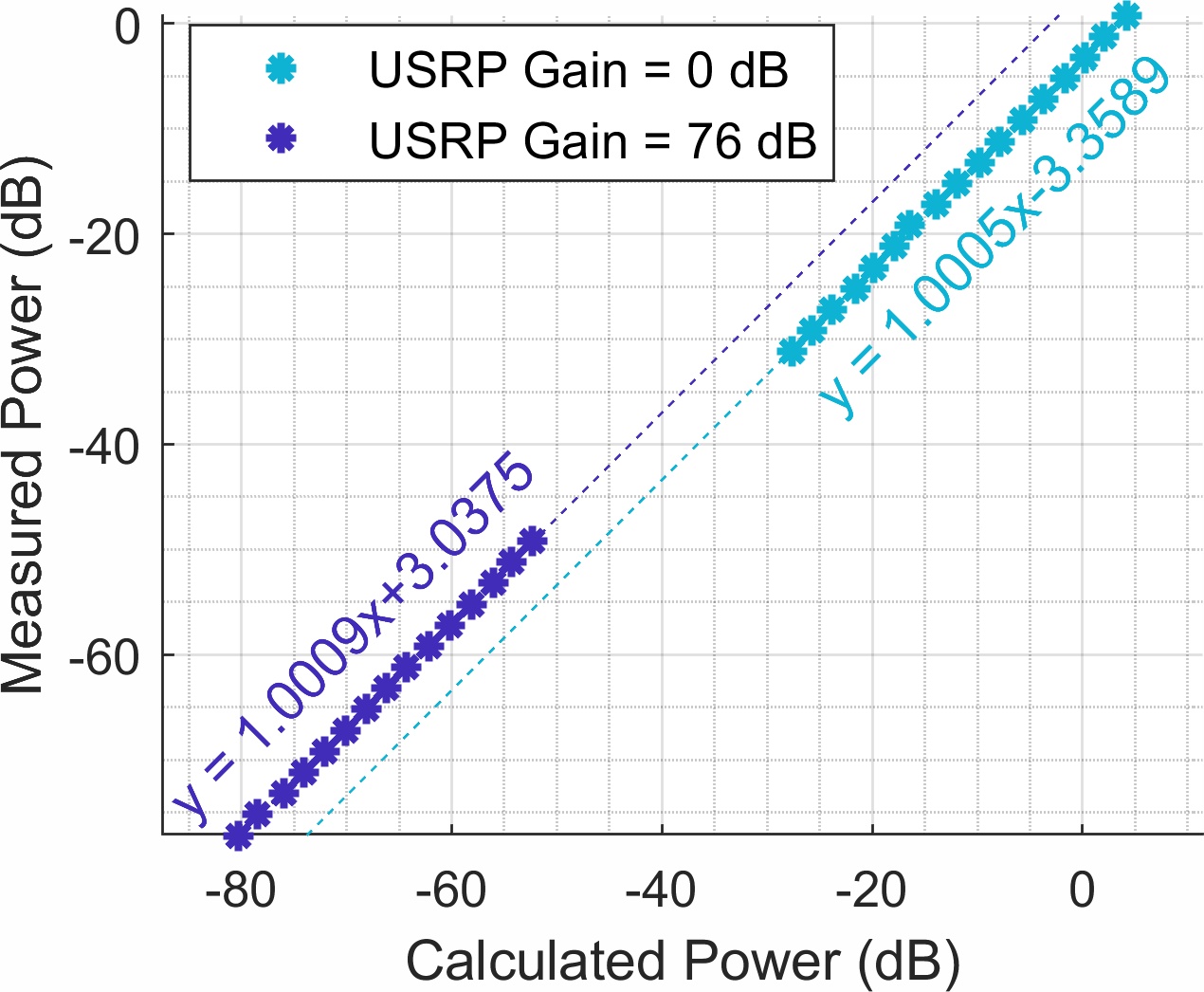}
         \caption{Calibration Curves}
     \end{subfigure}
     \begin{subfigure}{0.309\linewidth}
         \centering
         \includegraphics[width=1.0\linewidth]{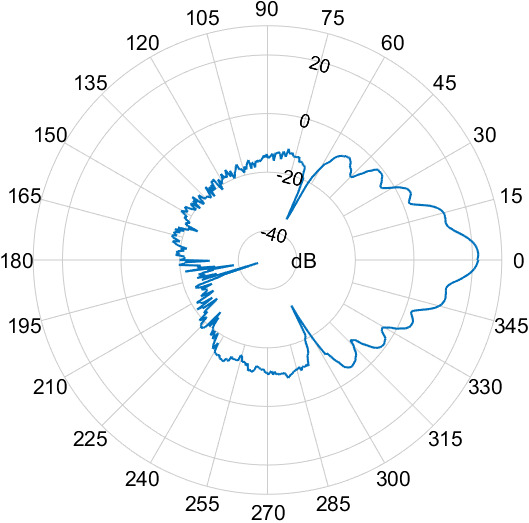}
         \caption{Normalized Antenna Pattern (Azimuth)}
         \label{F3b}
     \end{subfigure}
     \begin{subfigure}{0.309\linewidth}
         \centering
         \includegraphics[width=1.0\linewidth]{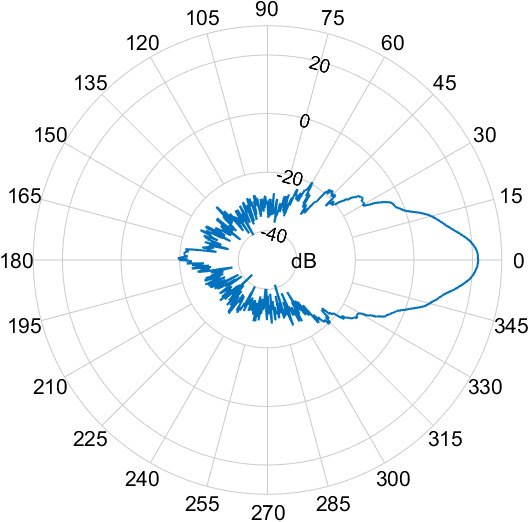}
         \caption{Normalized Antenna Pattern (Elevation)}
         \label{F3c}
     \end{subfigure}
     \vspace{-2mm}
     \caption{The calibration curves depicting measured power versus calculated power for USRP gains of \SI{0}{\deci\bel} and \SI{76}{\deci\bel} (a); and the normalized $2$D antenna patterns along the azimuth and elevation directions for the WR-$28$ horn antennas used in our channel sounder (b), (c).}
     \label{F3}
     \vspace{-4mm}
\end{figure*}

In this section, we outline the results of our evaluations on the collected dataset. First, in Fig.~\ref{F3}(a), we visualize the calibration curves, obtained according to the procedures outlined in Sec.~\ref{S3}, depicting the linear relationship between the power calculated (in dB) from USRP readings corresponding to the received power-delay profiles and the reference power levels (in dB). Also, obtained via empirical recordings of the WR-$28$ antenna's operational characteristics, Fig.~\ref{F3}(b) and Fig.~\ref{F3}(c) illustrate its $2$D radiation patterns along the azimuth and elevation directions, respectively: these enable us to compute the gains at specific locations along a route and at specific degrees of alignment, crucial for the analyses detailed next.

\begin{figure*} [t]
     \centering
     \begin{subfigure}{0.460\linewidth}
         \centering
         \includegraphics[width=1.0\linewidth]{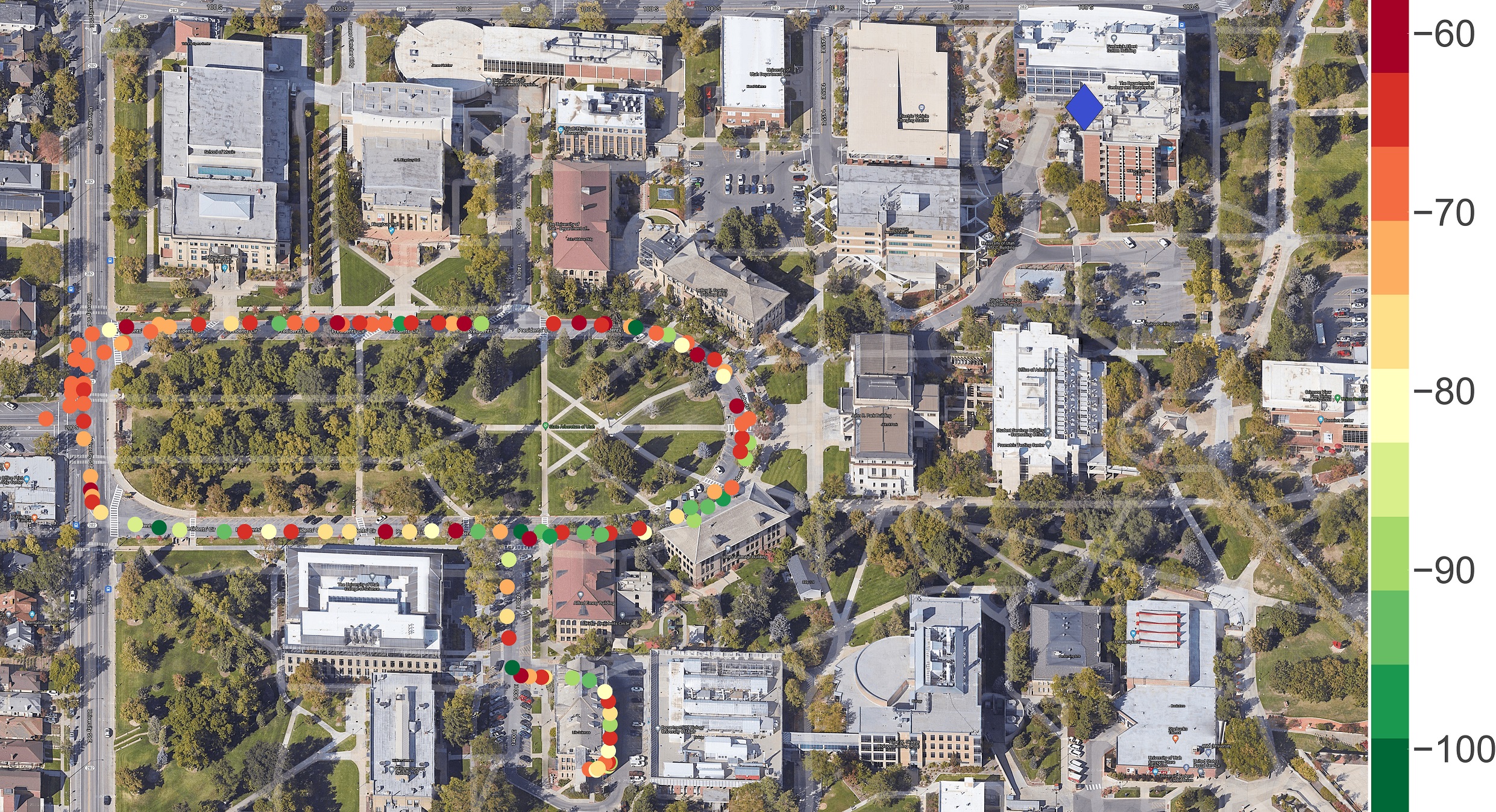}
         \caption{Rx Power: Urban Campus (President's Circle)}
         \label{F4a}
     \end{subfigure}
     \begin{subfigure}{0.2579\linewidth}
         \centering
         \includegraphics[width=1.0\linewidth]{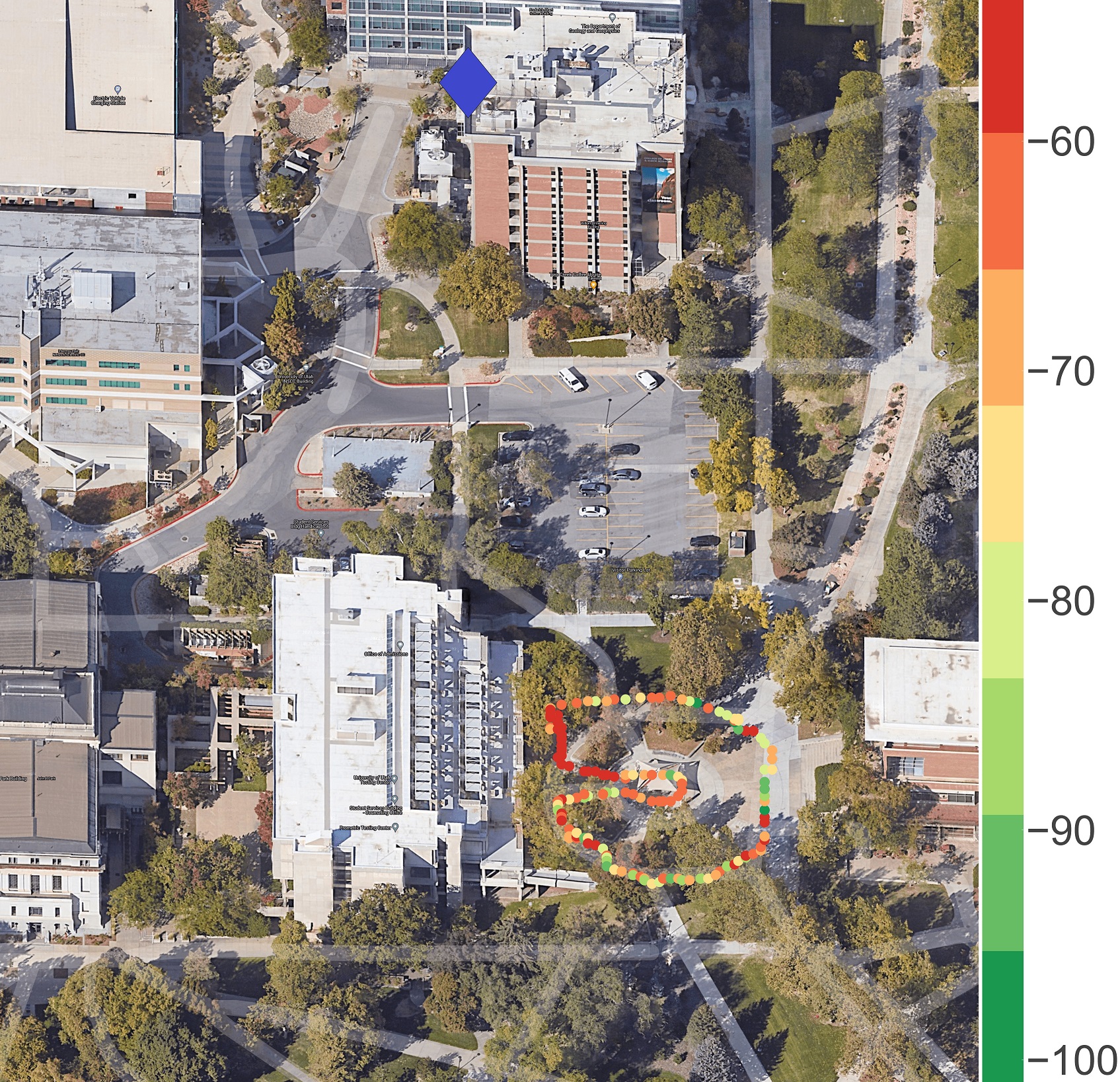}
         \caption{Rx Power: Urban Vegetation}
         \label{F4b}
     \end{subfigure}
     \begin{subfigure}{0.2579\linewidth}
         \centering
         \includegraphics[width=1.0\linewidth]{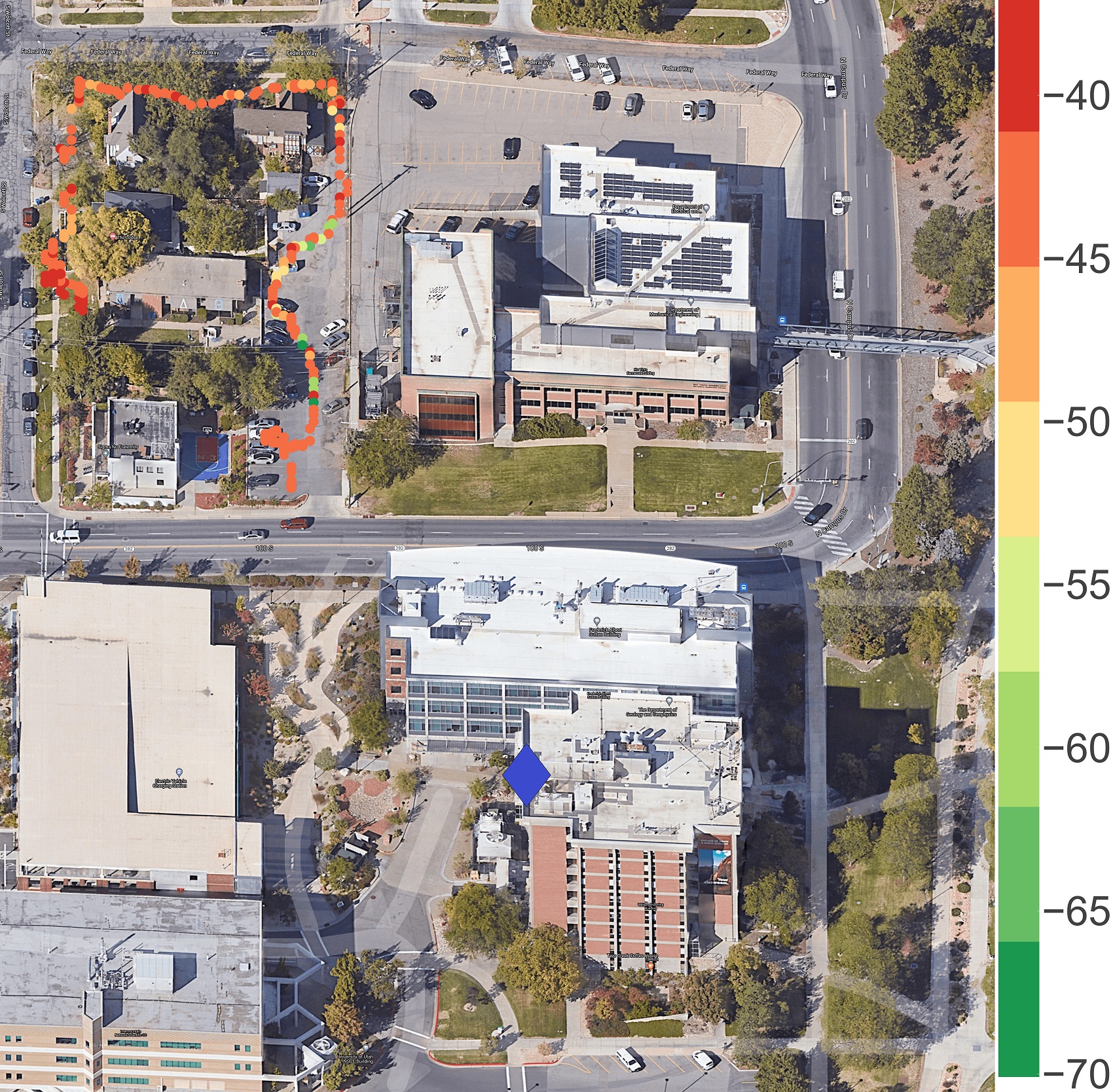}
         \caption{Rx Power: Suburban Neighborhood}
         \label{F4c}
     \end{subfigure}
     \vspace{-2mm}
     \caption{The received power values superimposed on a Google Hybrid map for urban campus (Rx on minivan), urban vegetation (Rx on cart), and suburban neighborhood (Rx on cart) routes, respectively. The heat-map color palette dots denote the Rx locations, while the purple diamond denotes the Tx location.}
     \label{F4}
     \vspace{-4mm}
\end{figure*}

\begin{figure*} [t]
     \centering
     \begin{subfigure}{0.48\linewidth}
         \centering
         \includegraphics[width=1.0\linewidth]{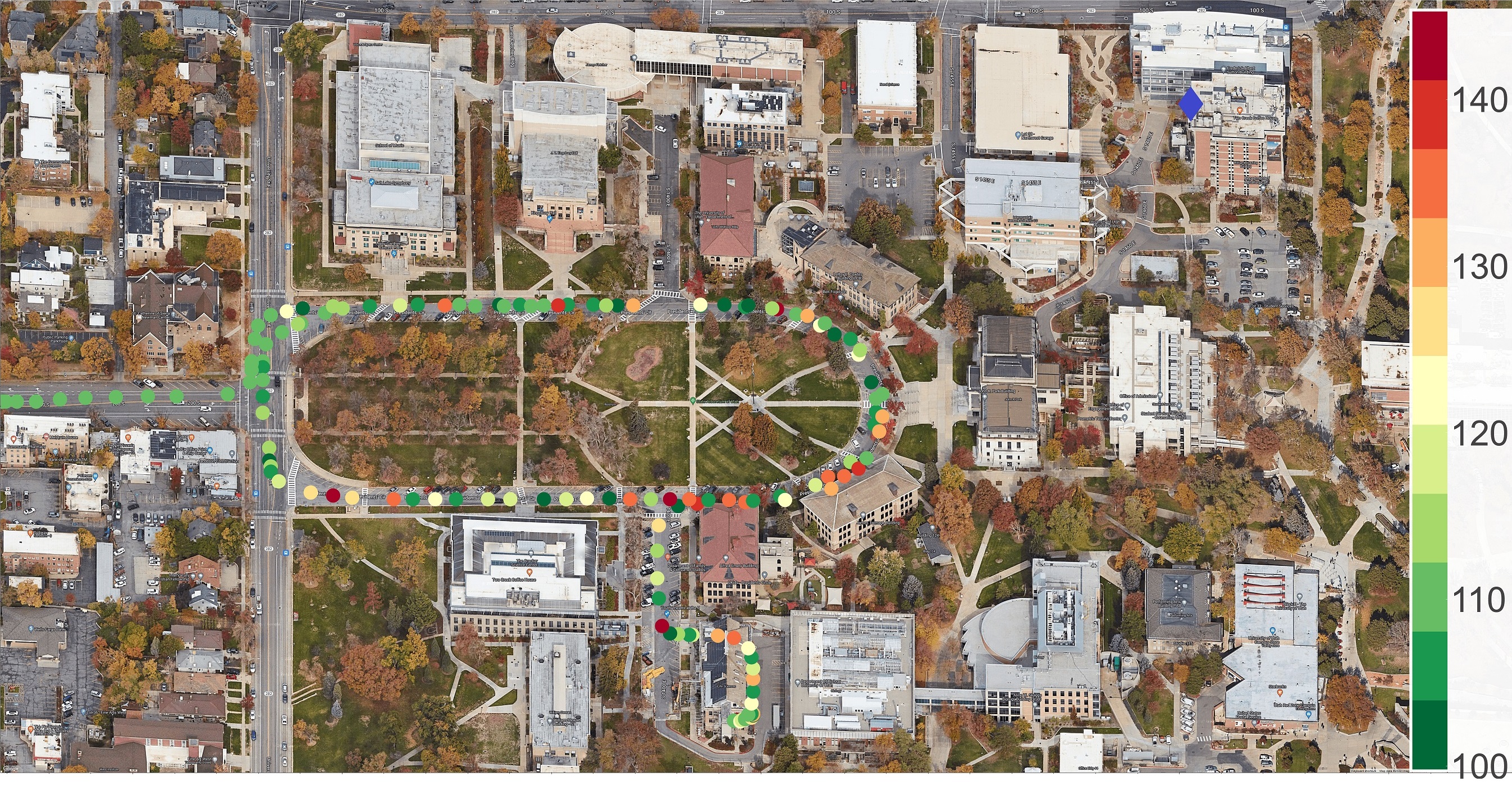}
         \caption{Pathloss: Urban Campus (President's Circle)}
         \label{F5a}
     \end{subfigure}
     \begin{subfigure}{0.245\linewidth}
         \centering
         \includegraphics[width=1.0\linewidth]{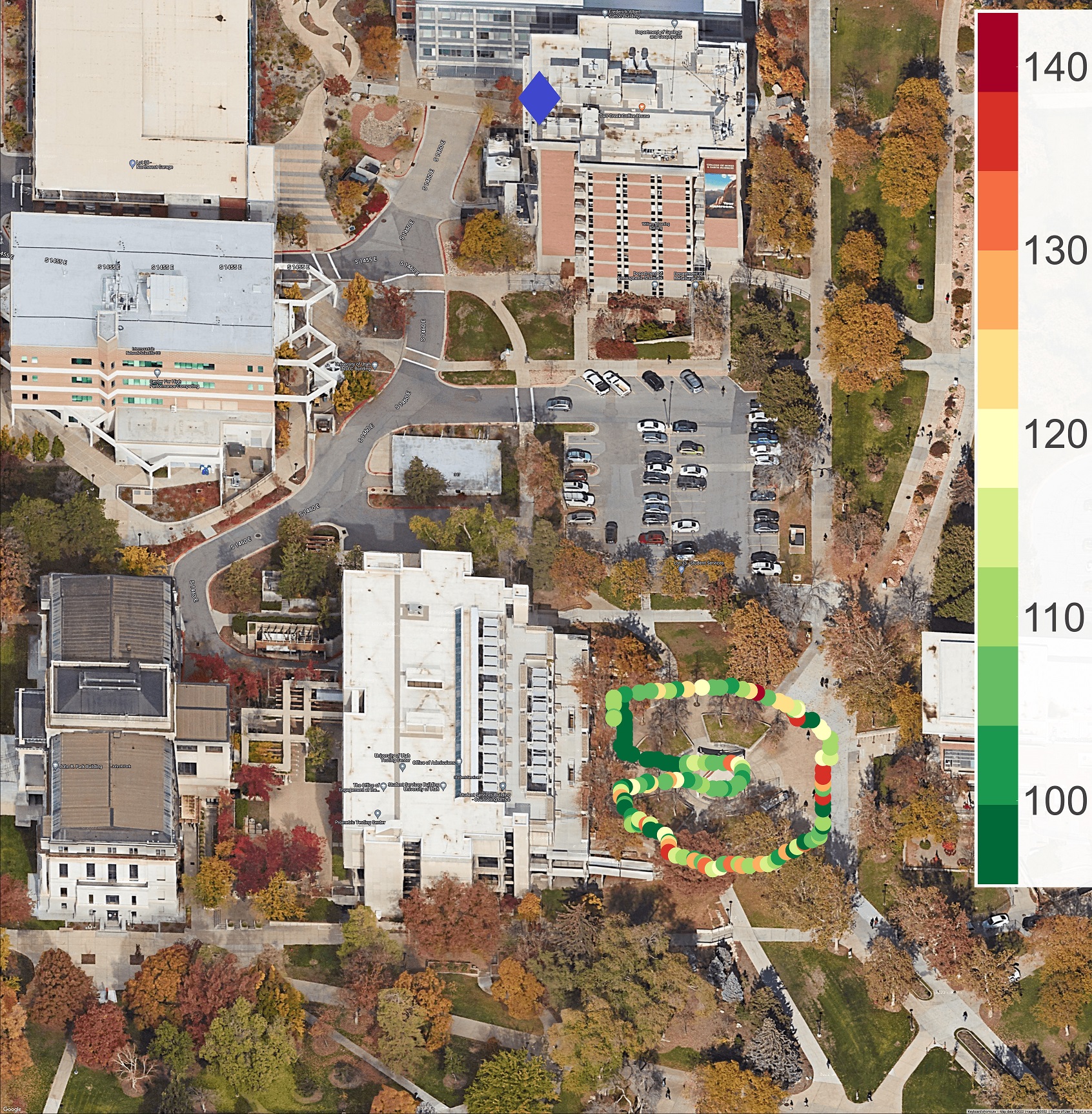}
         \caption{Pathloss: Urban Vegetation}
         \label{F5b}
     \end{subfigure}
     \begin{subfigure}{0.245\linewidth}
         \centering
         \includegraphics[width=1.0\linewidth]{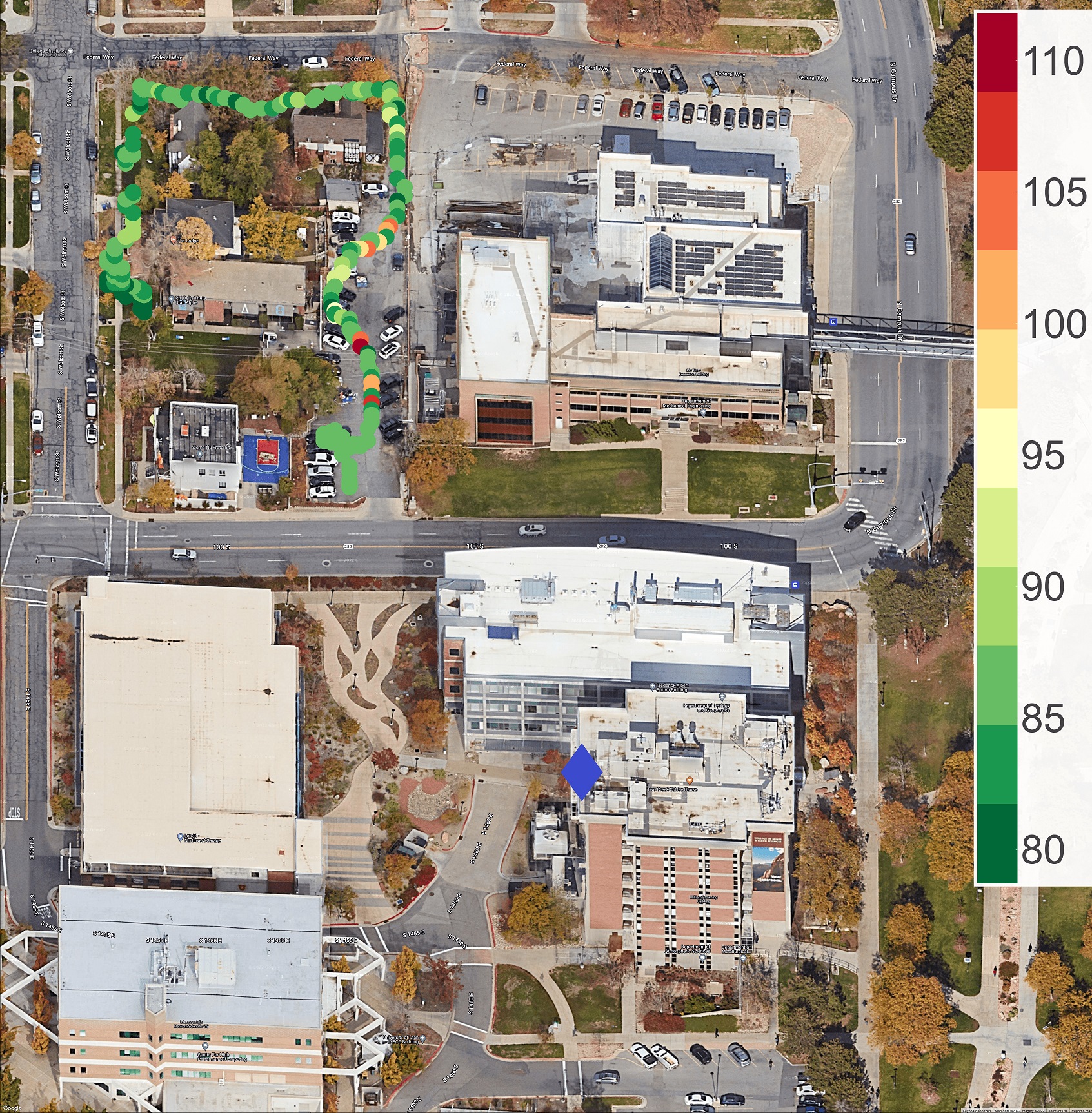}
         \caption{Pathloss: Suburban Neighborhood}
         \label{F5c}
     \end{subfigure}
     \vspace{-2mm}
     \caption{The pathloss values superimposed on a Google Hybrid map for urban campus (Rx on minivan), urban vegetation (Rx on cart), and suburban neighborhood (Rx on cart) routes, respectively. The heat-map color palette dots denote the Rx locations, while the purple diamond denotes the Tx location.}
     \label{F5}
     \vspace{-7mm}
\end{figure*}

\begin{figure*} [t]
     \centering
     \begin{subfigure}{0.496\linewidth}
         \centering
         \includegraphics[width=0.805\linewidth]{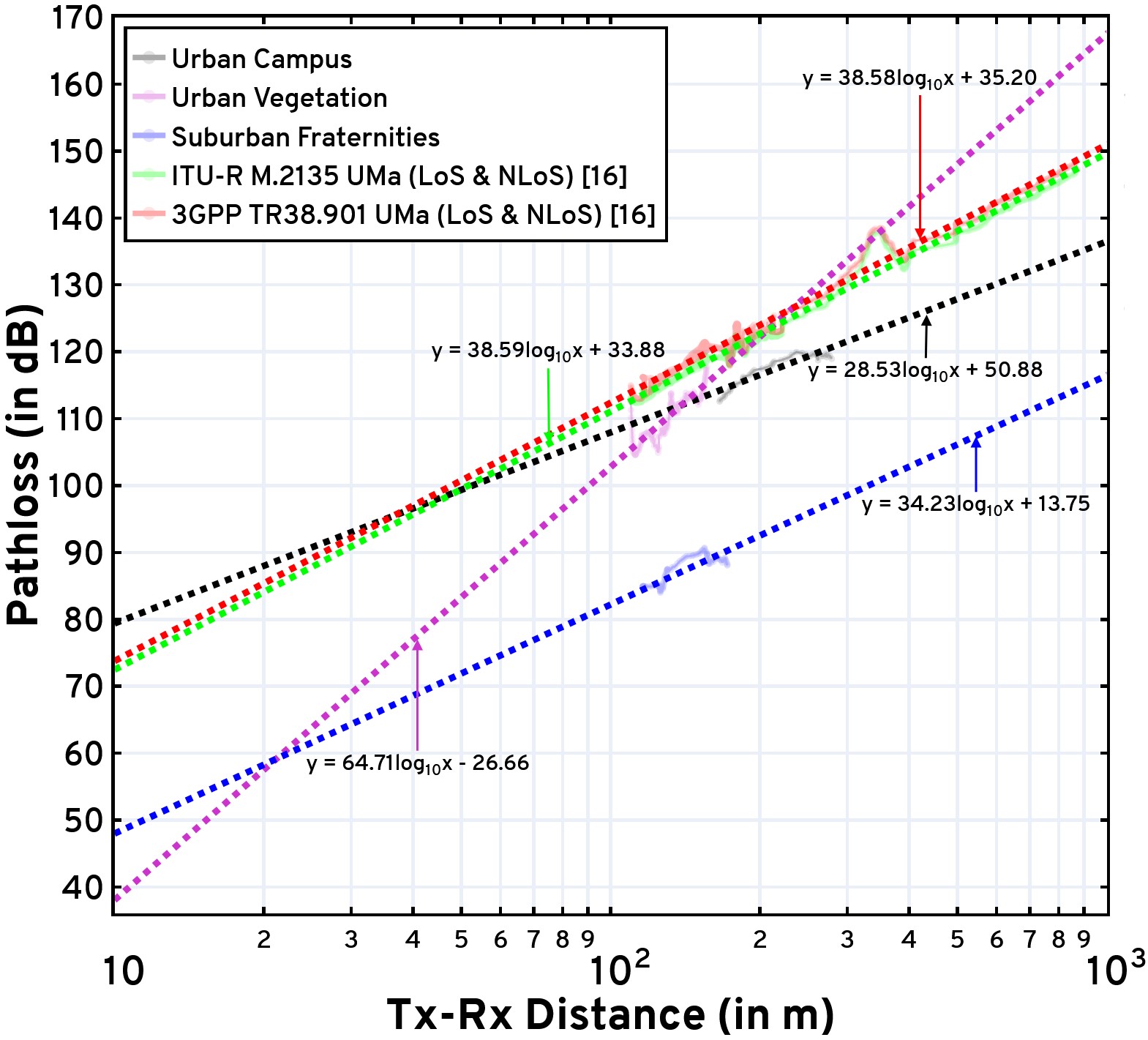}
         \vspace{-1mm}
         \caption{Pathloss vs Tx-Rx Distance}
         \label{F6a}
     \end{subfigure}
     \begin{subfigure}{0.496\linewidth}
         \centering
         \includegraphics[width=0.84\linewidth]{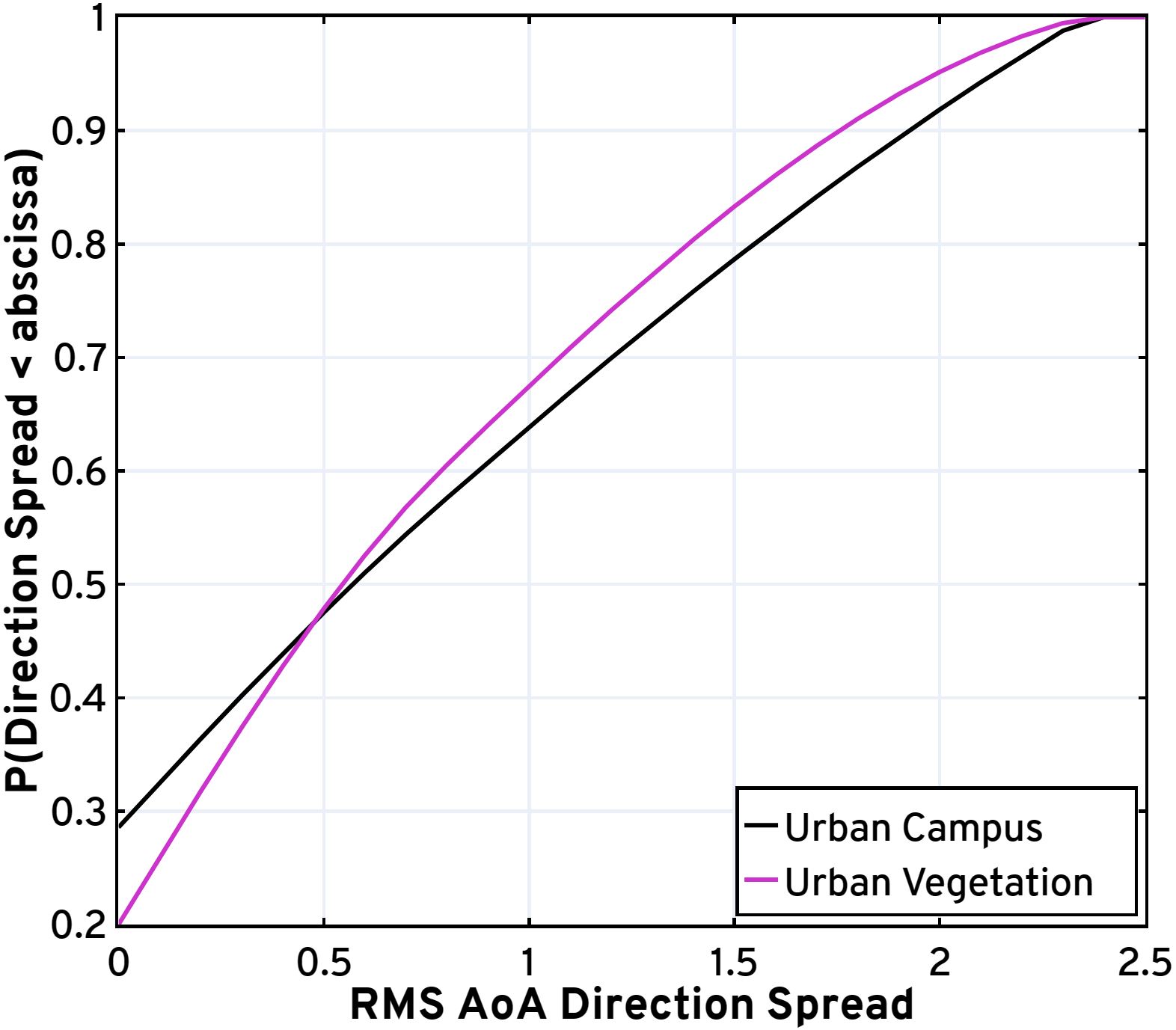}
         \vspace{-1mm}
         \caption{RMS AoA Direction Spread}
         \label{F6b}
     \end{subfigure}
     \vspace{-2mm}
     \caption{An illustration of the pathloss values computed from the various routes traversed during the measurement campaign compared against the $3$GPP TR$38.901$ and ITU-R M.$2135$ standards, with the dashed lines denoting the corresponding fitted models (a); and a plot depicting the cumulative distribution function of the RMS AoA direction-spread computed from the parameter estimates obtained via SAGE, for the urban-campus and urban-vegetation routes.}
     \label{F6}
     \vspace{-4mm}
\end{figure*}

\begin{figure*} [t]
     \centering
     \begin{subfigure}{0.496\linewidth}
         \centering
         \includegraphics[width=0.84\linewidth]{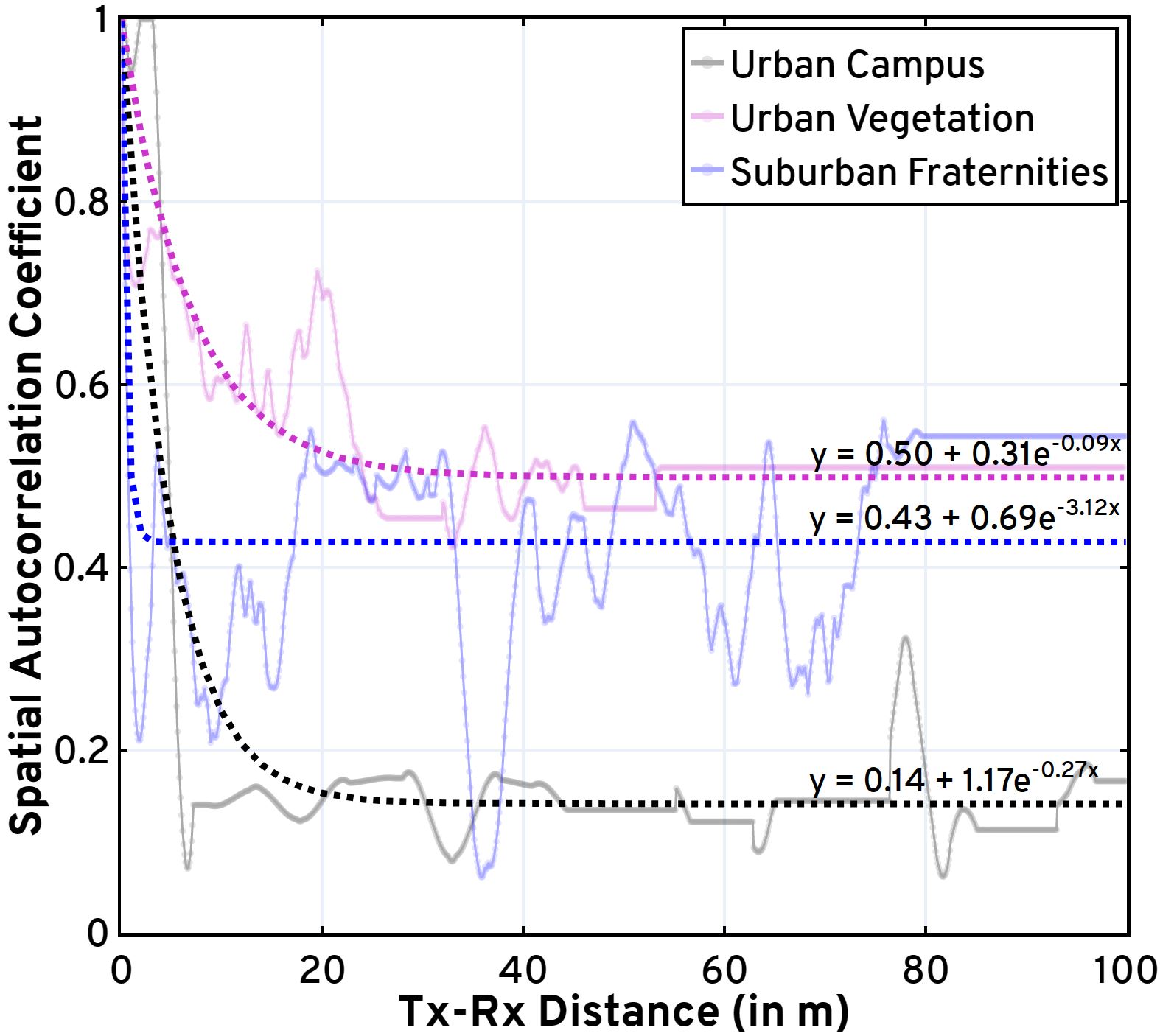}
         \vspace{-1mm}
         \caption{Spatial Consistency vs Tx-Rx Distance}
         \label{F7a}
     \end{subfigure}
     \begin{subfigure}{0.496\linewidth}
         \centering
         \includegraphics[width=0.83\linewidth]{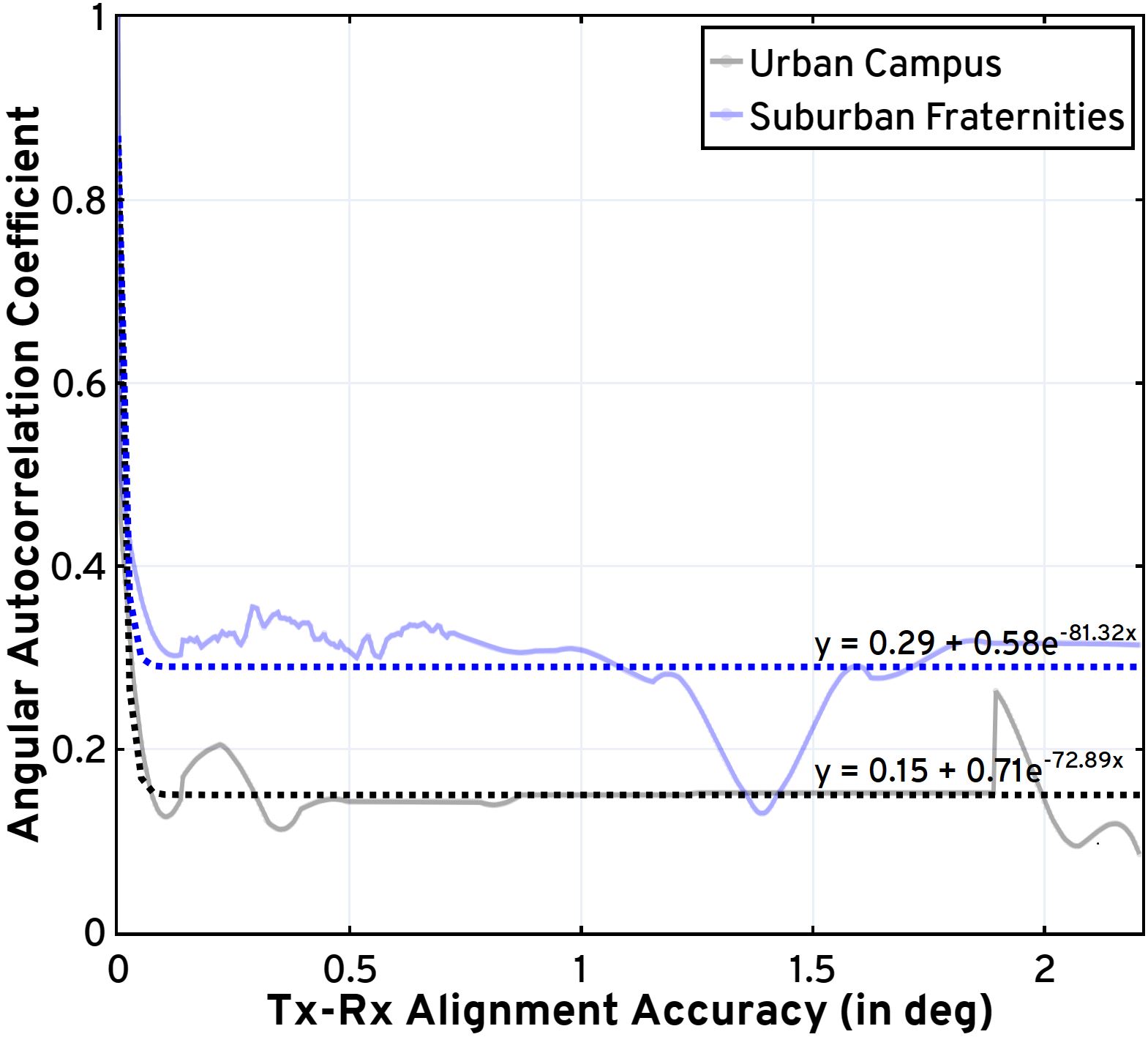}
         \vspace{-1mm}
         \caption{Spatial Consistency vs Tx-Rx Alignment Accuracy}
         \label{F7b}
     \end{subfigure}
     \vspace{-2mm}
     \caption{The plots depicting variations in the spatial/angular autocorrelation coefficient versus Tx-Rx distance (a) and Tx-Rx alignment accuracy (b).}
     \label{F7}
     \vspace{-6mm}
\end{figure*}

Fig.~\ref{F4} and Fig.~\ref{F5} depict the received power and pathloss heatmaps, superimposed on Google Hybrid maps of the routes traversed by the Rx onsite, with the Tx affixed on the rooftop of the William Browning building. Also, we compare the pathloss versus distance behavior of signals in our measurement campaign with large-scale urban macro-cellular pathloss standards ($3$GPP TR$38.901$ and ITU-R M$.2135$~\cite{MacCartneyModelsOverview}). These constitute both line-of-sight and non-line-of-sight models, with a Tx height of $h_{\text{Tx}}{\approx}$\SI{25}{\meter} and a Tx-Rx $2$D separation range of \SI{10}{\meter}${\leq}d_{2\text{D}}{\leq}$\SI{5000}{\meter}, which match the deployment specifications of our campaign, making them suitable candidates for empirical validations. As shown in Fig.~\ref{F6}(a), evaluating the pathlosses computed from our collected measurements against these standards for the urban campus (President's Circle), suburban neighborhood (S Walcott St), and urban vegetation (campus foliage) routes, we observe that while the $3$GPP and ITU-R standards model the empirical pathloss values for the urban campus route, they fail to accurately capture the pathloss vs distance behavior in suburban settings; additionally, these standards do not serve as good benchmarks for studying \SI{28}{\giga\hertz} propagation around foliage in urban environments.
\indent{Next}, to analyze the multipath propagation characteristics of \SI{28}{\giga\hertz} signals, we use the SAGE algorithm~\cite{SAGE} to extract the complex attenuation ($\alpha$), delay ($\tau$), Doppler shift ($\nu$), and angles-of-arrival ($\phi,\theta$) of the $M$ specular paths arriving at the Rx while traversing a particular route onsite. Solving for the exact high-resolution maximum likelihood estimate of this parameter vector ($\bm{\xi}_{i}{=}[\alpha_{i},\tau_{i},\nu_{i},\phi_{i},\theta_{i}]^{\intercal},i{=}1,2,{\dots},M$) directly involves prohibitively large computation times~\cite{SAGE}. Thus, the SAGE algorithm solves for an approximate estimate of $\bm{\xi}_{i}$ via iterative executions of the E-step which computes the expectation of the log-likelihood given the observations and the previous estimate, and the M-step which computes the current estimate by maximizing over the E-step result. This iterative execution occurs until convergence, i.e., the change in parameter values across consecutive iterations is smaller than a predefined threshold. Using these estimates, in Fig.~\ref{F6}(b) we plot the RMS direction-spread ($\sigma_{\Omega}$) of the received signals while traversing the urban campus and urban vegetation routes: these spread metrics are computed employing the procedures outlined in~\cite{Indoor60G}. From the estimated AoA azimuth and elevation values, we observe that the multipath components arriving at the Rx along the urban campus route demonstrate a relatively larger spread than those along the urban vegetation route.
\\\indent{Finally}, to study the spatial decoherence behavior of \SI{28}{\giga\hertz} signals in V$2$X scenarios, we evaluate the spatial/angular autocorrelation coefficient along specific Rx routes under the effects of Tx-Rx distance and alignment: this coefficient is computed according to the steps laid down in~\cite{MacCartneySpatialStatistics}, wherein for each Rx location change, the amplitude term in~\cite{MacCartneySpatialStatistics} constitutes the magnitude of all the power-delay profile components (along the excess delay axis) in a recorded measurement at that location, with the expectation taken over the ensemble. Fig.~\ref{F7}(a) illustrates these spatial consistency plots vis-\`{a}-vis distance, under perfect Tx-Rx alignment conditions, wherein exponential functions are fit to the computed coefficient values for the urban campus, suburban neighborhood, and urban vegetation routes. We note decreasing correlation trends between the recorded power-delay profile samples under increasing Tx-Rx separation. Similarly, keeping the Tx-Rx separation constant, Fig.~\ref{F7}(b) depicts the variation of the angular autocorrelation coefficient under increasing levels of misalignment between the Tx and Rx antennas, while traversing the urban campus and suburban neighborhood routes. We observe rapid decorrelation across the evaluated samples even at small amounts of misalignment, highlighting the characteristics of our directional WR-$28$ horn antennas, and illustrating the need for accurate beam-steering in mmWave V$2$X networks. In Figs.~\ref{F7}(a), (b), the channel does not get fully decorrelated since, in our \emph{beam-steered measurements}, the LoS component remains significant.
\vspace{-9mm}

% Concluding remarks and Future work
\balance
\section{Conclusion}\label{S5}
We discuss the design of a fully autonomous beam-steering platform coupled with a sliding correlator channel sounder, well-suited for mmWave V$2$X modeling. Corroborated onsite, this beam-steering system demonstrates superior performance vis-\`{a}-vis geo-positioning accuracy, alignment reliability, and tracking response times. Processing the power-delay profiles via custom noise elimination and thresholding heuristics, we perform pathloss evaluations against the $3$GPP TR$38.901$ and ITU-R M$.2135$ standards. Importantly, the continuous series of measurements facilitated by our design enables V$2$X spatial consistency studies vis-\`{a}-vis Tx-Rx distance and alignment.

% References (main.bib)
\balance
\bibliographystyle{IEEEtran}
\bibliography{IEEEabrv,main} 

\end{document}